\documentclass[conference]{IEEEtran}

\usepackage[usenames]{color}
\usepackage{balance}
\usepackage{amsmath}
\usepackage{amssymb}
\usepackage{multirow}
\usepackage{cite}
\usepackage{ifpdf}
\usepackage{epstopdf}
\usepackage{graphicx}
\usepackage{epsf}
\usepackage{hyperref}
\usepackage{fancyhdr}

\usepackage[printonlyused]{acronym}
\usepackage[letterpaper, left=1in, right=1in, bottom=1in, top=0.75in]{geometry}
\usepackage{nicefrac}
\newtheorem{theorem}{Theorem}

\usepackage{amsfonts}
\usepackage{amsmath}
\usepackage{mathrsfs}
\usepackage{balance}
\usepackage{cite}
\usepackage{subfigure}

\usepackage{listings}

\usepackage{color}
\usepackage{units}
\usepackage{nicefrac}
\usepackage[printonlyused]{acronym}
\usepackage[section] {placeins}
\usepackage{fancyhdr}
\definecolor{NewText}{rgb}{0.75,0,0.10}

\acrodef{BPP}[BPP]{binomial point process}
\acrodef{PPP}[PPP]{Poisson point process}
\acrodef{SNR}[SNR]{signal-to-noise ratio}
\acrodef{SINR}[SINR]{signal-to-interference-and-noise ratio}
\acrodef{PMF}[PMF]{probability mass function}
\acrodef{PDF}[PDF]{probability density function}
\acrodef{CDF}[CDF]{cumulative density function}
\acrodef{FHMA}[FHMA]{frequency-hopping multiple access}
\acrodef{MRC}[MRC]{maximal-ratio combining}
\acrodef{TCP}[TCP]{Thomas cluster process}

\begin{document}


\newcommand{\comment}[1]{\textcolor{red}{#1}}
\newcommand{\TBD}[1]{\textcolor{red}{TBD: #1}}

\hyphenation{multi-symbol}
\title{Outage Correlation in Finite and \\ Clustered Wireless Networks}
\author{\IEEEauthorblockN{ Salvatore Talarico,\IEEEauthorrefmark{1} Matthew C. Valenti\IEEEauthorrefmark{2}, and
Marco Di Renzo,\IEEEauthorrefmark{3} \\}
\IEEEauthorblockA{\IEEEauthorrefmark{1} Intel Mobile Communications Technology Ltd, Santa Clara, CA, USA. \\
\IEEEauthorrefmark{2}West Virginia University, Morgantown, WV, USA. \\
\IEEEauthorrefmark{3} CNRS-Centrale Supel\'ec-University of Paris-Sud XI, Paris, France.}
}

\date{}
\maketitle

\thispagestyle{empty}
\vspace{-1.20cm}

\begin{abstract}
Future wireless networks are expected to adopt many different network technologies and architectures that promise to greatly enhance data rate and provide ubiquitous coverage for end users, all while enabling higher spectral efficiency. These benefits come with an increased risk of co-channel interference and possible correlation in the aggregated interference, which impacts the communication performance. This paper introduces a mathematical framework to quantify the spatial correlation of the interference in finite and clustered wireless networks with signals subject to Rayleigh fading. Expressions are derived for the correlation coefficient  of  the  outage  probability  when  the  interferers are located according to some of the most common point processes used in the literature to model the spatial distribution of the devices: binomial point process (i.e., relay networks with fixed users), Poisson point process (i.e., heterogeneous cellular networks), and Thomas point process (i.e., device-to-device networks).
\end{abstract}

\section{Introduction} \label{Section:Intro}
Starting with a key result by Baccelli et al. \cite{baccelli:2006}, the communication theory community has embraced stochastic geometry over the past decade as a meaningful and tractable tool for the large-scale analysis and design of wireless networks.  See, for instance \cite{ganti:2009b,haenggi:2012b}
for comprehensive surveys of the application of stochastic geometry to wireless networks.  However, early applications of stochastic geometry focused on the asymptotic case of infinite networks with infinite numbers of interferers, typically drawn from a homogenous \ac{PPP}.  Actual networks have a finite number of interferers and occupy a finite area.  Recent work has focused on the important and realistic case of finite networks; see, for instance, \cite{torrieri:2012,valenti:2014,guo:2014}.

The basic application of stochastic geometry provides a snapshot view of the distribution of network interference taken at an arbitrary time and at an arbitrary location.  Such a view provides little insight into how performance seen by a particular user varies over time and distance, which is a critical consideration for the development of fundamental technologies whose performance depends on the spatio-temporal correlations, such as error-control coding, space-time signaling, and scheduling algorithms.   Recent work has focused on the analysis and quantification of \emph{interference correlation} \cite{ganti:2009,haenggi:2012,schilcher:2012,haenggi:2013d,haenggi:2013,zhong:2014,gong:2014,tanbourgi:2014,
wen:2015,Schilcher:2016,wen:2016,Ali:2017,Krishnan:2017a}.  Interference correlation arises because a common set of potential interferers are observed through similarly attenuated channels.  A pair of closely spaced receivers will receive similar interference even if the interference is received over fading channels that are decorrelated in space and time.  This is because the location of interferers are a common source of randomness, even if the fading channels that they are observed over are not common. Being able to accurately quantify the interference correlation is critical for many applications such as for cache-enabled cellular networks \cite{Krishnan:2017}, or to design and evaluate cooperative schemes \cite{Rajanna:2017} just to mention a few.

The goal of the present paper is to consider the issue of interference correlation in finite networks; i.e., networks that are finite in extent and in the number of users.  To the best of our knowledge, this paper is the first attempt to capture both finite network effects and interference correlation within a single unified analysis.  While related works use interference distribution, coverage probability, and rate distribution as metrics, we use the related metric of \emph{outage correlation}, which corresponds to the likelihood that a receiver at a particular location is in outage when a receiver at another nearby location is also in outage.  While our work is initially applied to interference drawn from a homogenous point process, we also consider, as an extension, the outage correlation when the interferers are drawn from a clustered process.  For ease of exposition, it is assumed that the fading is Rayleigh, but a similar procedure can be used to extend the analysis to Nakagami-m fading. The analysis builds on a method \cite{valenti:2014} for obtaining the spatially averaged outage probability in finite networks, which starts with a closed-form expression for the outage probability conditioned on a fixed topology (but averaged over the fading) and then averages over the position of the interferers and the number of interferers.  Expressions for the outage correlation coefficient are provided in closed form up to a one-dimensional integral that could be pre-computed and tabulated, much like the Gaussian-Q function.

\section{Related Work}

The spatio-temporal interference correlation is considered in \cite{ganti:2009} for an infinite network of interferers drawn from a \ac{PPP} under the assumption of an Aloha MAC protocol.  In \cite{haenggi:2012}, Haenggi shows that the interference correlation produces a significant diversity loss in a \ac{PPP} network with multi-antenna receivers. In \cite{schilcher:2012}, Schilcher et al. focused on the temporal correlation of the interference and derived equations for the correlation coefficient of the interference power in two consecutive time slots in a variety of scenarios. In \cite{haenggi:2013d}, using the {\em diversity polynomial} defined in \cite{haenggi:2012}, which captures the temporal interference correlation, it is shown that the joint success probability of multiple transmissions can be expressed in closed-form for a \ac{PPP} network, and in this case there is no retransmission diversity for simple retransmission schemes. In \cite{haenggi:2013}, a stochastic geometry-based mathematical framework is proposed to analyze the mean time required to connect to a nearest neighbor, namely local delay, for both a highly mobile as well as a static Poisson network. In \cite{zhong:2014}, Zhong et al. studied the effect of temporal interference correlation on \ac{FHMA} and Aloha MAC protocols. They derived closed-form expressions for the mean and variance of the local delay for the two MAC protocols when the nodes are drawn from a \ac{PPP}, and they evaluated the channel access probability that minimizes the mean local delay. In \cite{gong:2014}, Gong et al. evaluated the temporal correlation of the interference and outage in mobile Poisson networks for several mobility models.

More recently, Tanbourgi et al. have introduced an analytical framework in \cite{tanbourgi:2014} to evaluate the performance of \ac{MRC} in the presence of spatially correlated interference across antennas in an infinite \ac{PPP} network.
The effect of interference correlation is quantified in \cite{wen:2015} for a heterogeneous cellular network under the assumption of a K-tier Poisson infinite network. A variant of the Campbell-Mecke theorem is proposed in \cite{Schilcher:2016}, which is used to evaluate the link outage in Nakagami fading for Poisson networks to quantify the temporal interference correlation. In \cite{wen:2016}, the authors provided an analytical framework to quantify the interference correlation in non-Poisson networks including Mattern cluster networks and second-order cluster networks, and show that the decrease of the mean number of mobiles in each cluster or the increase in the radius of each cluster mitigates the interference correlation. In \cite{Ali:2017}, the authors studied the effect of spatial interference correlation on secure communication in the downlink of a Poisson cellular network where UEs have full-duplex jamming capability when an eavesdropper lies near the UE under consideration. In \cite{Krishnan:2017a}, the authors studied the effect of interference correlation for cellular network and quantified the effect of handoff in terms of coverage probability in a stochastic geometry fashion.

\section{Network Model} \label{Sec:Network_Model}

\begin{figure}[!t]
\centering
\includegraphics[width=7.0cm]{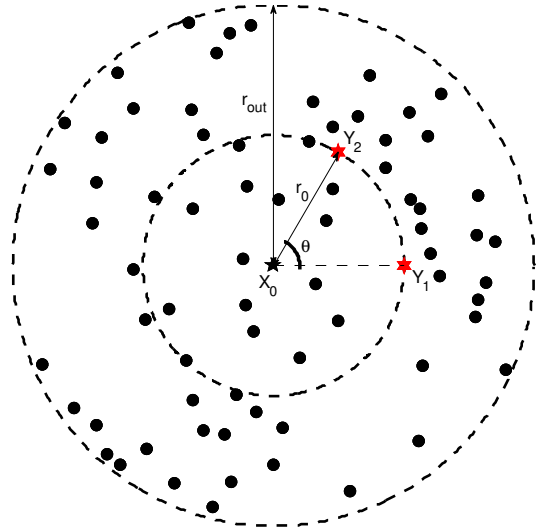}
\vspace{-0.25cm}
\caption{Illustration of a typical network. $X_0$ is the reference transmitter, $Y_1$ and $Y_2$ are two receivers whose correlation we seek, and the black dots are potential sources of interference.}  \label{Fig.1}
\vspace{-0.5cm}
\end{figure}

Consider a circular network $\mathcal A$ with area $|\mathcal A|$.
A reference transmitter $X_0$ is located at the center of the network, and its signal is received at two locations, $Y_1$ and $Y_2$.  For ease of exposition, assume that both receivers are at the same distance $r_0$ from the transmitter, though the analysis can be extended to handle receivers at different distances.  Representing each location $Y_j$ by a complex number, $Y_1 = r_0$ and $Y_2 = r_0 e^{j \theta}$; i.e., $\theta$ is the angle between the two receive locations on the radius-$r_0$ circle.  The network furthermore contains $M$ potentially interfering transmitters $\{X_i,...,X_M\}$. The number of interferers $M$ could be either fixed or random.
Fig. \ref{Fig.1} shows a typical network topology, where the black dots indicate the position of the interfering transmitters, the five-point star in the center is the reference transmitter, and the two red six point stars are the two receivers.

The power of transmitter $X_i$'s signal at location $Y_j$ is
$
  \rho_{i,j}
   =
  P_i g_{i,j}  d_{i,j}^{-\alpha}
$
where $P_i$ is the transmit power associated with $X_i$, $g_{i,j}$ is the power gain due to fading, $d_{i,j} = ||X_i-Y_j||$ is the distance between transmitter $X_i$ and location $Y_j$, and $\alpha$ is the path-loss exponent. Note that the path-loss model includes a singularity as $d_{i,j} \rightarrow 0$, however as we show, the singularity is inconsequential in the analysis.
Independent Rayleigh fading is assumed, in which case the $\{g_{i,j}\}$ are i.i.d. unit-mean exponential, though the analysis could be easily extended to handle Nakagami-m fading.

The instantaneous \ac{SINR} at the reference location $Y_j$ is
\begin{eqnarray}
   \gamma_j
   & = &
   \frac{ \rho_{0,j} }{ \displaystyle {\mathcal N} + \sum_{i=1}^{M} I_i \rho_{i,j} }
=
   \frac{ g_{0,j}  }{ \displaystyle \mathsf{SNR}^{-1} + \sum_{i=1}^M I_i g_{i,j} r_{i,j}^{-\alpha} }
   \label{Equation:SINR1}
\end{eqnarray}
where $\mathcal N$ is the noise power, $I_i$ is a Bernoulli variable with $\mathbb P[I_i=1] = p_i$, $\mathsf{SNR} = P_0 r_0^{-\alpha}/\mathcal{N}$ is the \ac{SNR} at the receiver, and $r_{i,j} = d_{i,j}/r_0$ is the normalized distance between the $i^{th}$ transmitter and the $j^{th}$ receive location.  The last step assumes that  $P_i = P_0$ for all $i \in \{1,...,M\}$, and $p_i$ is a \textit{random-access probability} $p_i$ can be used to model an Aloha-like protocol.

\section{Analytical Framework} \label{Sec:Analysis}
Let $\boldsymbol r = \{ r_{i,j}, i \in \{1,...,M\}, j \in \{1,2\} \}$ represent the network topology.  Given $\boldsymbol r$ and $M$, the outage probability may be found at either receive location:
\begin{eqnarray*}
  \epsilon_j ( \boldsymbol r  )
  & = &
  \mathbb P [ \gamma_j \leq \beta | \boldsymbol r, M ]
  =
  F_{\gamma_j}( \beta | \boldsymbol r, M ).
\end{eqnarray*}
By conditioning on the network topology and the number of interferers, the outage probability can be evaluated as
\vspace{-0.05cm}
\begin{eqnarray}
\epsilon_j(\boldsymbol r)
 \hspace{-0.05cm}=\hspace{-0.05cm}
1 \hspace{-0.05cm}- \hspace{-0.05cm}\exp \hspace{-0.05cm}\left(\hspace{-0.05cm}- \frac{\beta}{\mathsf{SNR} }\hspace{-0.05cm}\right) \hspace{-0.05cm}\prod_{i=1}^M \hspace{-0.05cm} \left(  \hspace{-0.05cm} 1 \hspace{-0.05cm}- \hspace{-0.05cm}p_i \hspace{-0.05cm}+ \hspace{-0.05cm}\frac{p_i r_{i,j}^{\alpha}}{\beta +  r_{i,j}^{\alpha} } \right).
\label{Equation:RayleighConditional}
\end{eqnarray}

Because of the randomness of $\boldsymbol r$ and $M$, $\epsilon_j\left(\boldsymbol r\right)$ is a random variable.  The outage probabilities $\epsilon_1\left(\boldsymbol r\right)$ and $\epsilon_2\left(\boldsymbol r\right)$ at the two locations constitute a pair of random variables, and as such, they may be correlated. The spatial correlation of the outages is here quantified through the correlation coefficient of the outage probability $\zeta$ at receiver $Y_1$ and $Y_2$. Since the outage probabilities at receiver $Y_1$ and $Y_2$ are statistically equivalent,
\begin{eqnarray}
\zeta[Y_1,Y_2]
\hspace{-0.25 cm}& = & \hspace{-0.25 cm}
 \frac{\mathbb E_{M,\boldsymbol r}\left[\epsilon_1\left( \boldsymbol r\right) \epsilon_2\left( \boldsymbol r\right)\right] - \mathbb E_{M,\boldsymbol r}^2\left[\epsilon_j\left( \boldsymbol r\right)\right] }{\mathbb E_{M,\boldsymbol r}\left[ \epsilon_j^2\left( \boldsymbol r\right) \right] - \mathbb E_{M,\boldsymbol r}^2\left[ \epsilon_j\left( \boldsymbol r\right) \right]}.
\label{correlation_coefficient}
\end{eqnarray}

The analysis introduced here follows from and extends the methodology introduced in \cite{valenti:2014}, which evaluates moments in (\ref{correlation_coefficient}) in two steps: 1) the number of interferers $M$ is fixed and then the average is found with respect to $\boldsymbol r$, and 2) the number of interferers is allowed to be random (e.g,. as in a PPP), and the average found with respect to $M$. Specifically, the {\em first moment} of the outage probability conditioned on $M$ is
\begin{eqnarray}
   \mathbb E_{\boldsymbol r }\left[\epsilon_j\left( \boldsymbol r \right) | M \right]
  \hspace{-0.25 cm} & = &\hspace{-0.25 cm}
   \int
   \epsilon_j\left( \boldsymbol r \right)
   {f}_\mathbf{\boldsymbol r }\left( \boldsymbol r | M \right)
    \,{\rm d} \mathbf{\boldsymbol r}.
    \label{eq:FirstMoment}
\end{eqnarray}

The {\em second moment} of the outage probability conditioned on $M$ is
\begin{eqnarray}
   \mathbb E_{\boldsymbol r }\left[\epsilon_j^2\left( \boldsymbol r \right)  | M \right]
   \hspace{-0.25 cm} & = & \hspace{-0.25 cm}
   \int
   \epsilon_j^2\left( \boldsymbol r \right)
   {f}_\mathbf{\boldsymbol r}\left( \boldsymbol r | M \right)
    \,{\rm d} \mathbf{\boldsymbol r}.
    \label{eq:SecondMoment}
\end{eqnarray}
The {\em first joint moment} of the outage probability conditioned on $M$ is
\begin{eqnarray}
   \mathbb E_{\boldsymbol r}\left[\epsilon_1\left( \boldsymbol r \right) \epsilon_2 \left( \boldsymbol r \right)| M \right] \hspace{-0.25 cm} & =& \hspace{-0.25 cm}
\int \hspace{-0.05 cm}{ \epsilon}_1 \hspace{-0.05 cm} \left[ \boldsymbol r \right] \hspace{-0.05 cm} { \epsilon}_2 \hspace{-0.05 cm} \left[ \boldsymbol r \right] \hspace{-0.05 cm} {f}_\mathbf{\boldsymbol r }\left( \boldsymbol r | M \right) \hspace{-0.05 cm} \,{\rm d} \mathbf{r}.
    \label{eq:JointFirstMoment}
\end{eqnarray}

When the number of interferers is random, the overall spatially averaged first, second and first joint moment of the outage probability can be obtained by averaging with respect to $M$. In particular, it is possible to use the following equality
  \begin{eqnarray}
  \mathbb E_{M,\boldsymbol r}\hspace{-0.05 cm} \left[\epsilon_i^x\left( \hspace{-0.05 cm} \boldsymbol r \hspace{-0.05 cm} \right) \hspace{-0.05 cm} \epsilon_j^y\left( \hspace{-0.05 cm} \boldsymbol r \hspace{-0.05 cm} \right)\right]
   \hspace{-0.05 cm} \hspace{-0.05 cm}= \hspace{-0.05 cm} \hspace{-0.05 cm}
   \sum_{ m=0}^{ \infty } \hspace{-0.05 cm} p_{M}[m] \hspace{-0.05 cm} E_{\boldsymbol r} \hspace{-0.05 cm} \left[\epsilon_i^x\left( \boldsymbol r \right)\hspace{-0.05 cm} \epsilon_j^y\left( \boldsymbol r \right) | m \right]
   \label{eq:JointFirstMomentAveraged}
\end{eqnarray}
where $p_{M}[m]$ is the \ac{PMF} of $M$.

\subsection{Binomial Point Process}
\label{subsection:BPP_correlation}

 Assume that $p_i=p_0$ for all $i$, and that there are a fixed number of $M$ interferers that are independently and uniformly distributed (i.u.d) over $\mathcal A$. Thus, the interferers are drawn from a BPP of intensity $\lambda = M/|\mathcal A|$.

\begin{theorem}
The spatially averaged first moment of the outage probability is
\begin{eqnarray}
  \hspace{-0.4cm } \mathbb E_{\boldsymbol r }\left[\epsilon_j\left( \boldsymbol r \right) \hspace{-0.05 cm}| \hspace{-0.05 cm} M \right]
  \hspace{-0.35cm }& =& \hspace{-0.35cm } 1 \hspace{-0.1cm }-\hspace{-0.1cm } \exp \hspace{-0.05 cm} \left( \hspace{-0.05 cm} \frac{\beta}{\mathsf{SNR}} \right) \hspace{-0.1cm } \left[ 1 \hspace{-0.05 cm}-\hspace{-0.05 cm}p \hspace{-0.05 cm}+\hspace{-0.05 cm}
  p \mathcal{T} \left( r_\mathsf{out} , r_\mathsf{0} \right)
   \right]^M  \label{eq:FirstMomentFinal}
\end{eqnarray}
where the function $\mathcal{T} \left( r_\mathsf{out} , r_\mathsf{0} \right)$ is given by (\ref{eq:T_func}).

\underline{Proof:} The derivation is provided in Appendix \ref{App:A}.
\end{theorem}

\begin{theorem}
The spatially averaged second moment of the outage probability is
\begin{eqnarray}
\mathbb E_{\boldsymbol r}\left[\epsilon_j^2 \left( \boldsymbol r \right) | M \right]
\hspace{-0.25 cm}&= & \hspace{-0.25 cm} 2 E_{\boldsymbol r}\left[\epsilon_j\left( \boldsymbol r \right) | M \right]-1 \nonumber \\
\hspace{-0.25 cm} & & \hspace{-0.25 cm} + \exp\left(- \frac{2 \beta }{\mathsf{SNR}}\right) \mathcal{S}\left( r_\mathsf{out} , r_{\mathsf{0}} \right)^M
\label{eq:SecondMomentFinal}
\end{eqnarray}
where the function $\mathcal{S}\left( r_\mathsf{out} , r_{\mathsf{0}} \right)$ is given by (\ref{integral_S(x,y)}).

\underline{Proof:} The derivation is provided in Appendix \ref{App:B}.
\end{theorem}

\begin{theorem}
The spatially averaged joint first moment of the outage probability is
\begin{eqnarray}
\hspace{-0.5 cm}\mathbb E_{\boldsymbol r}\left[\epsilon_1\left( \boldsymbol r \right) \epsilon_2\left( \boldsymbol r \right)| M \right]
\hspace{-0.25 cm}&= & \hspace{-0.25 cm} 2 E_{\boldsymbol r}\left[\epsilon_j\left( \boldsymbol r \right) | M \right]-1 \nonumber \\
\hspace{-0.75 cm} & & \hspace{-1 cm}
+ \exp\left(- \frac{2 \beta }{\mathsf{SNR}}\right) \mathcal{W}\left( r_\mathsf{out} , r_{\mathsf{0}},\theta \right)^M
\label{eq:JointFirstMomentFinal}
\end{eqnarray}
where the function $\mathcal{W}\left( r_\mathsf{out} , r_{\mathsf{0}},\theta \right)$ is given by (\ref{eq:W(x,y,z)}).

\underline{Proof:} The derivation is provided in Appendix \ref{App:C}.
\end{theorem}

Finally, by substituting (\ref{eq:FirstMomentFinal}), (\ref{eq:SecondMomentFinal}) and (\ref{eq:JointFirstMomentFinal}) in (\ref{correlation_coefficient}), the correlation coefficient can be evaluated.

\subsection{Poisson Point Process}
\label{subsection:PPP_correlation}

Suppose that the interferers are drawn from a \ac{PPP} with intensity $\lambda$ on the plane $\mathbb R^2$.
The number of interferers $M$ in $\mathcal A$ is Poisson with mean $\mathbb E[M] = \lambda |\mathcal A|$.  It follows that the \ac{PMF} of the number of interferers within $\mathcal A$ is given by \cite{ganti:2009b}
\begin{eqnarray}
p_{M}[m]
\hspace{-0.3 cm} & = & \hspace{-0.3 cm} \frac{(\lambda |\mathcal A|)^{m}}{m!} \exp\left( -\lambda |\mathcal A| \right), \; \; \; \mbox{for $m \geq 0$}.
\label{pmf_Poisson}
\end{eqnarray}

Set $x=0$ and $y=1$ in (\ref{eq:JointFirstMomentAveraged}), and substitute (\ref{pmf_Poisson}), and (\ref{eq:FirstMomentFinal}) within it. By applying the power-series representation of the exponential function, the spatially averaged first moment of the outage probability is
\begin{eqnarray}
 \hspace{-0.3 cm} \mathbb E_{M,\boldsymbol r}\left[\epsilon_i\left( \boldsymbol r \right)\right] \hspace{-0.3 cm}&= & \hspace{-0.3 cm}
  1 - \exp\left\{-\frac{\beta}{\mathsf{SNR}} - 2 \lambda p   \left[\frac{|\mathcal A|}{2}
  -  \frac{\pi r_{\mathsf{0}}^2}{\beta} \right. \right. \nonumber \\
  \hspace{-0.6 cm}& & \hspace{-0.3 cm}
  \left. \left.
  \times  \Psi \hspace{-0.05 cm} \left( \hspace{-0.05 cm} \frac{r_\mathsf{out}-r_{\mathsf{0}}}{r_{\mathsf{0}}} \hspace{-0.05 cm} \right) \hspace{-0.05 cm} + \hspace{-0.05 cm} r_{\mathsf{0}}^2 \mathcal{Z}_1\left( r_\mathsf{out} , r_{\mathsf{0}} \hspace{-0.05 cm} \right) \hspace{-0.05 cm}\right]\hspace{-0.05 cm}
  \right\} \label{cdf_PPP_expectation}
\end{eqnarray}
where the function $\Psi \left(x\right)$ is given by \ref{Ihyp}, and the function $\mathcal{Z}_i\left( x, y\right)$ is given by \ref{eq:Z_i}.

Set $x=0$ and $y=2$ in (\ref{eq:JointFirstMomentAveraged}), and substitute (\ref{pmf_Poisson}) and (\ref{eq:SecondMomentFinal}) within it. By applying again the power-series representation of the exponential function, the spatially averaged second moment of the outage probability is
\begin{eqnarray}
 \mathbb E_{M,\boldsymbol r }\left[\epsilon_i^2 \left( \boldsymbol r \right) \right]\hspace{-0.25 cm}&= & \hspace{-0.25 cm}
  2 \mathbb E_{M,\boldsymbol r}\left[\epsilon_i\left( \boldsymbol r \right)\right] -1
   \nonumber \\
  \hspace{-0.25 cm}& & \hspace{-2 cm}
  + \exp\left\{-\frac{2 \beta }{\mathsf{SNR}}
  - \lambda |\mathcal A| \left[ 1+ \mathcal{S}\left( r_\mathsf{out} , r_{\mathsf{0}} \right) \right]\right\}. \label{cdf_PPP_variance}
\end{eqnarray}

Set $x=1$ and $y=1$ in (\ref{eq:JointFirstMomentAveraged}), and substitute (\ref{pmf_Poisson}) and (\ref{eq:JointFirstMomentFinal}) within it. By using the power-series representation of the exponential function, the spatially averaged first joint moment of the outage probability is
\begin{eqnarray}
\mathbb E_{M , \boldsymbol r}\left[\epsilon_1\left( \boldsymbol r \right) \epsilon_2\left( \boldsymbol r \right) \right]
\hspace{-0.25 cm}&= & \hspace{-0.25 cm} 2 E_{M, \boldsymbol r}\left[\epsilon_i \left( \boldsymbol r \right) \right]-1
 \nonumber \\
\hspace{-0.45 cm}& & \hspace{-3 cm}
+ \hspace{-0.05 cm} \exp\left\{\hspace{-0.05 cm} - \frac{2 \beta}{\mathsf{SNR}}
\hspace{-0.05 cm} - \hspace{-0.05 cm} \lambda |\mathcal A| \left[ 1\hspace{-0.05 cm}+\hspace{-0.05 cm} \mathcal{W}\left( r_\mathsf{out} , r_{\mathsf{0}},\theta \right) \right] \right\}.
\label{cdf_PPP_covariance}
\end{eqnarray}

Finally, by substituting (\ref{cdf_PPP_expectation}), (\ref{cdf_PPP_variance}) and (\ref{cdf_PPP_covariance}) in (\ref{correlation_coefficient}), the correlation coefficient can be evaluated.

\vspace{-0.25cm}

\subsection{Thomas Cluster Process}
\label{subsection:TPP_correlation}

Assume the location of the interferers are modeled by a \ac{TCP}, which is a special case of a Poisson cluster process, and is generated by a set of offspring points independently and identically distributed (i.i.d.) around each point of a parent \ac{PPP} \cite{ganti2}. In particular, the locations of parent points are  modeled  as  a  homogenous  \ac{PPP} with intensity $\lambda_\text{parent}$ around  which  offspring  points  are  distributed  according to a symmetric normal distribution with  variance $\sigma^2$. The number of offspring points per parent point is fixed and is Poisson with intensity $\lambda'$.  Let $\boldsymbol \nu$ be the set of relative distances from a receiver to the center of the cluster\footnote{For conciseness, we provide here only analytical expressions for intra-cluster interferences, assuming the cluster is centered at the reference transmitter location.}. For any $\nu \in \boldsymbol \nu$, the \ac{PDF} of $\boldsymbol r$ conditioned  over $\nu$ is \cite{Afshang:2016}
\begin{eqnarray}
   {f}_{r}( {r}|\nu ) &=&  \frac{r}{\sigma^2} \exp\left( - \frac{r^2 + \nu^2}{2 \sigma^2} \right) \mathcal{I}_0\left( \frac{r \nu }{\sigma^2} \right) \label{pdf_tx_cluster}
\end{eqnarray}
where
$\mathcal{I}_0\left( \cdot \right)$is the modified Bessel function with order zero.

\begin{theorem}
The spatially averaged first moment of the outage probability is
\begin{eqnarray}
\hspace{-0.1cm } \mathbb E_{M, \boldsymbol r }\left[\epsilon_j\left( \boldsymbol r \right)\right]
  \hspace{-0.35cm } & =& \hspace{-0.35cm } 1 \hspace{-0.05cm }- \hspace{-0.05cm } \exp\left( \hspace{-0.05cm }- \frac{\beta}{\mathsf{SNR}}\hspace{-0.05cm } -\hspace{-0.05cm } \lambda' p \hspace{-0.05cm } + \hspace{-0.05cm } \lambda' p \mathcal{V} \left( \sigma \right) \hspace{-0.05 cm} \hspace{-0.05cm }\right)
\label{eq:JointFirstMomentFinalTCP}
\end{eqnarray}
where the function $\mathcal{V} \left( \sigma \right) $ is given by (\ref{eq:V(sigma)}).

\underline{Proof:} The derivation is provided in Appendix \ref{App:D}.
\end{theorem}
\begin{theorem}
The spatially averaged second moment of the outage probability is
\begin{eqnarray}
\mathbb E_{M,\boldsymbol r }\left[\epsilon_i^2 \left( \boldsymbol r \right) \right]\hspace{-0.25 cm}&= & \hspace{-0.25 cm} 2 \mathbb E_{M,\boldsymbol r}\left[\epsilon_i\left( \boldsymbol r \right)\right] -1
   \nonumber \\
  & & \hspace{-1 cm}
  + \exp\left\{-\frac{2 \beta }{\mathsf{SNR}}
  - \lambda' \left[ 1+ \mathcal{Q}\left( \sigma \right) \right]\right\}. \label{eq:JointSecondMomentFinalTCP}
\end{eqnarray}
where the function $\mathcal{Q} \left( \sigma \right) $ is given by (\ref{integral_Q(sigma)}).

\underline{Proof:} The derivation is provided in Appendix \ref{App:E}.
\end{theorem}

\begin{theorem}
The spatially averaged joint first moment of the outage probability is
\begin{eqnarray}
\mathbb E_{M , \boldsymbol r}\left[\epsilon_1\left( \boldsymbol r \right) \epsilon_2\left( \boldsymbol r \right) \right]
\hspace{-0.25 cm}&= & 2 E_{M, \boldsymbol r}\left[\epsilon_i \left( \boldsymbol r \right) \right]-1
 \nonumber \\
\hspace{-0.45 cm}& & \hspace{-2 cm}
+ \exp\left\{- \frac{2 \beta}{\mathsf{SNR}}
- \lambda' \left[ 1+ \mathcal{D}\left( \sigma, \theta \right) \right]\right\} \label{eq:JointFirstMomentFinalTCP2}
\end{eqnarray}
where the function $\mathcal{D} \left( \sigma, \theta \right) $ is given by (\ref{integral_D(sigma)}).

\underline{Proof:} The derivation is provided in Appendix \ref{App:F}.
\end{theorem}


\section{Numerical Results} \label{Sec:Results}

 In this section, we illustrate the efficacy of the proposed analysis by showing the outage correlation for a network with interferers drawn from \ac{BPP}, \ac{PPP}, and \ac{TCP} processes.  Fig. \ref{Fig.2} shows the results for \ac{BPP} and \ac{PPP} processes, while Fig. \ref{Fig.3} shows results for the \ac{TCP} process.
 In all results shown in this section, the path loss exponent is $\alpha=3.5$, $\mathsf{SNR}=10$ dB, and $\beta=0$ dB. In both  Fig. \ref{Fig.2}, and  Fig. \ref{Fig.3}, the lines are generated analytically using the methodologies in this paper, while the markers are obtained by simulation, which involved generating $10^5$ network topologies $\mathbf r$ based on the spatial model used and evaluating for each of them the conditional outage probability by (\ref{Equation:RayleighConditional}). Both figures show a good agreement between analytical and simulated results, and confirms the correctness of the analysis.

 Fig. \ref{Fig.2} shows the spatially averaged correlation coefficient of the outage probability as function of $\theta$. The network is here assumed to be circular with radius $r_{\mathsf{out}}=1$, and both receivers are at distance $r_{0}=0.25$ from the reference transmitter located at the center.
Results are shown for both the \ac{BPP} and \ac{PPP} cases and for three values of $\lambda p$.
 For the \ac{BPP} networks, the number of interferers is fixed at $M=50$, implying that $\lambda = M /| \mathcal A| = M /\pi$, so that $p=\{0.1,0.5,1\}$ achieves $\lambda p = \{5/\pi,25/\pi,50/\pi\}$, respectively.
As shown, spatial correlation increases as the two receive locations get closer together (decreasing $\theta$).  Moreover, sparser networks are more correlated than denser networks.  This is because an outage in a sparse network is likely due to an interferer that is close to one of the receivers, and if the receivers are sufficiently close together the very same interferer is likely to cause an outage at the other receiver. As the number of interferers increases, there are more dominant interferers (which are closer to the receivers) and each one  of them has an  independently  fading  channel, which decorrelates the events. Furthermore, Fig. \ref{Fig.2} shows  that a \ac{PPP} experiences a higher correlation than a \ac{BPP}. This can be attributed to the random number of interferers in a PPP and the non-negligible chance that a given instance of a PPP is sparse and hence highly correlated.


\begin{figure*}[t!]
\begin{eqnarray}
   \mathbb E_{\boldsymbol r }\left[\epsilon_j^2\left( \boldsymbol r \right)  | M \right]
   =
1 - 2 \exp\left(-\frac{\beta}{\mathsf{SNR}}\right) \left[  1 - p + p \mathcal{T} \left( r_\mathsf{out} , r_\mathsf{0} \right) \right]^M  + \exp\left(-2 \frac{\beta }{\mathsf{SNR}}\right)   \mathcal{S} \left( r_\mathsf{out} , r_\mathsf{0} \right) ^M.
\label{cdfwithintegral_square}
\end{eqnarray}
\vspace{-0.25cm}
{\hrulefill}
\end{figure*}

\begin{figure*}[t!]
\begin{eqnarray}
\mathbb E_{\boldsymbol r}\left[\epsilon_1\hspace{-0.05cm}\left( \boldsymbol r \right) \epsilon_2\hspace{-0.05cm}\left( \boldsymbol d \right)\hspace{-0.05cm}| \hspace{-0.05cm}M \right]
\hspace{-0.3cm} & = & \hspace{-0.3cm} 1 \hspace{-0.05cm}- \hspace{-0.05cm} \exp \hspace{-0.05cm} \left( -\frac{\beta }{\mathsf{SNR}}\hspace{-0.05cm} \right)  \left[  1 \hspace{-0.05cm}- \hspace{-0.05cm}p \hspace{-0.05cm}+ p
R\left( 0\right)
  \right]^M
 - \exp\left( -\frac{\beta }{\mathsf{SNR}}\right) \left[ 1  - p  +  p
R \left( \theta\right) \right]^M  + \exp\left(-\frac{2 \beta }{\mathsf{SNR}}\right) \nonumber \\
\hspace{-0.3cm} & & \hspace{-0.3cm} \times \left[   \int \hspace{-0.3cm} \int
\left( 1- p +
\frac{ p r^{\alpha}\left(\rho,\phi, 0 \right) }{\beta + r^{\alpha}\left(\rho,\phi, 0 \right) }
\right) \left( 1- p +
\frac{ p r^{\alpha}\left(\rho,\phi, \theta \right) }{\beta + r^{\alpha}\left(\rho,\phi, \theta \right) }
\right) f_{\rho}\left( \rho\right) f_{\phi}\left( \phi\right) \,{\rm d} \rho \,{\rm d} \phi \right]^M.
\label{eq:JointFirstMomentFinal_Derivations_1}
\end{eqnarray}
\vspace{-0.25cm}
{\hrulefill}
\end{figure*}


\begin{figure*}[t!]
\begin{eqnarray}
   \mathbb E_{\boldsymbol r }\left[\epsilon_j^2\left( \boldsymbol r \right)  | m \right]
   =
1 - 2 \exp\left(-\frac{\beta}{\mathsf{SNR}}\right) \left[  1 - p + p \mathcal{V} \left( \sigma \right) \right]^m  + \exp\left(-2 \frac{\beta }{\mathsf{SNR}}\right)   \mathcal{Q} \left( \sigma \right) ^m.
\label{cdfwithintegral_square_TCP}
\end{eqnarray}
\vspace{-0.3cm}
{\hrulefill}
\end{figure*}

\begin{figure}
\centering
\includegraphics[width=8.0cm]{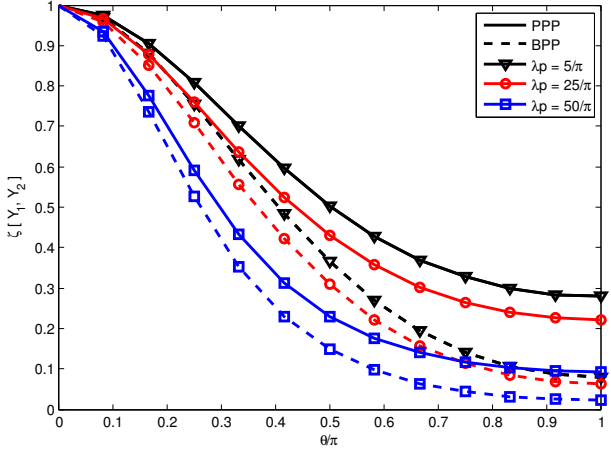}
\vspace{-0.35cm}
\caption{Correlation coefficient of the outage probability as function of $\theta$ when interferers are drawn from both a \ac{BPP} and a \ac{PPP}.}  \label{Fig.2}
\vspace{-0.05cm}
\end{figure}

\begin{figure}[t]
\centering
\includegraphics[width=8.0cm]{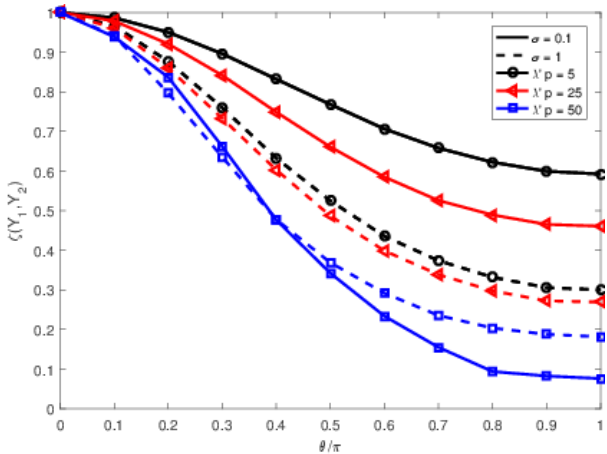}
\vspace{-0.35cm}
\caption{Correlation coefficient of the outage probability as function of $\theta$ when interferers are drawn from \ac{TCP}.}  \label{Fig.3}
\vspace{-0.35cm}
\end{figure}

 Fig. \ref{Fig.3} shows the spatially averaged correlation coefficient when interferers are drawn from a \ac{TCP}. Results are shown as function of $\theta$ for three values of $\lambda' p$ and two values of $\sigma$. Fig. \ref{Fig.3} shows that as expected, a more compact cluster (a cluster characterized by a lower $\sigma$) experiences higher correlation, but as this becomes denser the correlation decreases faster, since there are more dominant interferers.

\vspace{-0.25cm}

\section{Conclusion} \label{Section:Conclusion}
Even when interference is received through independently faded channels, the outage probabilities will be correlated at different receiver locations due to the fact that the location of the interferers is a common source of randomness.   This paper has provided numerical expressions for the correlation coefficient corresponding to the spatial outage probability of a finite wireless network.  We found that sparser networks have a higher spatial correlation, and that PPP networks have a higher correlation than BPP because of the non-negligible probability that a PPP network will be relatively sparse.  The spatial correlation in a TCP is smaller when the offspring points are more highly dispersed. We anticipate that the correlation coefficient quantified in this paper will be useful for many modern applications, including cache-enabled cellular networks \cite{Krishnan:2017} and cooperative schemes \cite{Rajanna:2017}.



\appendices
\section{} \label{App:A}
This appendix provides details leading to (\ref{eq:FirstMomentFinal}). For this scenario the PDF of $\{r_{i,j}\}$ is \cite{khalid:2013}
\begin{eqnarray}
   {f}_{r}( {r})
   \hspace{-0.10cm} = \hspace{-0.10cm}
   \begin{cases}
\frac{2 \pi {r} r_\mathsf{0}^2}{|{\mathcal A}|}
      &
      \hspace{-0.3cm} \mbox{\small for $ 0 \leq {r} \leq \tau_1 $} \\
\frac{2 {r} r_\mathsf{0}^2 }{|{\mathcal A}|}   \arccos \left(  \frac{r_\mathsf{0}^2 {r}^2 + r_\mathsf{0}^2 - r_\mathsf{out}^2}{2 r_\mathsf{0}^2 {r}} \right) &
       \hspace{-0.3cm} \mbox{\small for $  \tau_1 \leq  r \leq \tau_2$}
   \end{cases} \label{pdf_tx_in_center}
\end{eqnarray}

where $\displaystyle \tau_1= \frac{r_\mathsf{out}- r_\mathsf{0}}{r_\mathsf{0}}$ and $\displaystyle \tau_2=\frac{r_\mathsf{out}+r_\mathsf{0}}{r_\mathsf{0}}$.

The spatially averaged first moment of the outage probability is derived by substituting
(\ref{Equation:RayleighConditional}) and (\ref{pdf_tx_in_center}) in (\ref{eq:FirstMoment}).
Since the $M$ interferers are i.u.d. over $\mathcal A$,
\begin{eqnarray}
   \mathbb E_{\boldsymbol r }\left[\epsilon_j\left( \boldsymbol r \right)\hspace{-0.05cm} | \hspace{-0.05cm} M \right]
\hspace{-0.1cm} &= &\hspace{-0.1cm} 1 \hspace{-0.05cm} - \hspace{-0.05cm} \exp \hspace{-0.05cm} \left( \hspace{-0.05cm} - \hspace{-0.05cm} \frac{\beta }{\mathsf{SNR}} \hspace{-0.05cm} \right) \nonumber \\ && \hspace{-0.1cm} \times \left[ 1\hspace{-0.05cm}-\hspace{-0.05cm}p \hspace{-0.05cm}+ \hspace{-0.05cm} p \mathcal{T} \left( r_\mathsf{out} , r_\mathsf{0} \right)
    \right]^{M} \hspace{-0.1cm}
\label{eq:firstmoment3}
\end{eqnarray}
where
\begin{eqnarray}
\hspace{-0.4cm}\mathcal{T} \left( r_\mathsf{out} , r_\mathsf{0} \right)  \hspace{-0.3cm}  &=&  \hspace{-0.3cm} \int\frac{r^{\alpha+1}}{\beta +r^{\alpha}} f_r\left( r \right) \,{\rm d} r \nonumber \\
 \hspace{-0.7cm}  &=&  \hspace{-0.3cm}
\frac{2 \pi r_\mathsf{0}^2 }{|{\mathcal A}| \beta } \Psi \hspace{-0.05cm} \hspace{-0.05cm}\left( \hspace{-0.05cm} \hspace{-0.05cm} \frac{r_\mathsf{out} - r_\mathsf{0}}{r_\mathsf{0}} \hspace{-0.05cm} \right)
   \hspace{-0.05cm} \hspace{-0.05cm}+  \hspace{-0.05cm} \hspace{-0.05cm}\frac{2 r_\mathsf{0}^2 }{|{\mathcal A}| } \mathcal{Z}_i \hspace{-0.05cm} \left( \hspace{-0.05cm} r_\mathsf{out} , r_\mathsf{0} \hspace{-0.05cm} \right) \label{eq:T_func}
\end{eqnarray}

\begin{eqnarray}
\hspace{-0.45cm}\Psi\left( x \right) \hspace{-0.3cm} & =& \hspace{-0.35cm}  \int_{0}^{x} \frac{r^{\alpha+1}}{1+r^{\alpha}\beta^{-1}} \,{\rm d} r \label{eq:Phi1} \\
\hspace{-0.45cm} \mathcal{Z}_i \left( x, y \right)
\hspace{-0.3cm} \hspace{-0.05cm} \hspace{-0.05cm} &= & \hspace{-0.05cm} \hspace{-0.05cm} \hspace{-0.35cm}
\int_{x-y}^{ x+y } \hspace{-0.05cm} \hspace{-0.15cm}
 \frac{r^{i\alpha+1}}{\left(\beta \hspace{-0.05cm} + \hspace{-0.05cm} r^{\alpha}\right)^i } \hspace{-0.05cm} \arccos \hspace{-0.05cm} \left( \hspace{-0.05cm} \frac{y^2 r^2 \hspace{-0.05cm} + \hspace{-0.05cm} y^2 \hspace{-0.05cm} \hspace{-0.05cm} - \hspace{-0.05cm} x^2}{2 y^2 r} \hspace{-0.05cm} \right) \hspace{-0.1cm} \hspace{-0.05cm} \,{\rm d} r. \label{eq:Z_i}
\end{eqnarray}

By performing the change of variable $\nu = (r/y)^\alpha$,
\begin{eqnarray}
\Psi(x)
\hspace{-0.2cm} & = & \hspace{-0.2cm}
\frac{x^{\alpha+2}}{\alpha}
\int_{ 0 }^{ 1 }
\nu^{\frac{2}{\alpha}}
\left( 1 +
\frac{x^\alpha}{\beta} \nu
\right)^{-1}
d \nu \nonumber \\
\hspace{-0.2cm} & = & \hspace{-0.2cm}
\left( \hspace{-0.05cm}
\frac{x^{\alpha+2}}{2+\alpha}
\hspace{-0.05cm} \right)\hspace{-0.05cm}
\hspace{-0.01cm}_{2}F_{1} \hspace{-0.05cm}
\left(\hspace{-0.05cm}
\left[ 1, \frac{2}{\alpha} \hspace{-0.05cm} + \hspace{-0.05cm} 1 \right]\hspace{-0.05cm};\hspace{-0.05cm}
\frac{2}{\alpha} \hspace{-0.05cm}+ \hspace{-0.05cm} 2;
- \hspace{-0.05cm}\frac{x^\alpha}{\beta }\hspace{-0.05cm}
\right) \label{Ihyp}
\end{eqnarray}

where $_{2}F_{1}$ is  the Gauss hypergeometric function.

\section{} \label{App:B}
This appendix provides details leading to (\ref{eq:SecondMomentFinal}). The spatially averaged second moment of the outage probability is derived by substituting (\ref{Equation:RayleighConditional}) and (\ref{pdf_tx_in_center}) in (\ref{eq:SecondMoment}). Since the $M$ interferers are i.u.d. over $\mathcal A$, a few algebraic manipulations yields (\ref{cdfwithintegral_square}), where the function $\mathcal{T} \left( r_\mathsf{out} , r_\mathsf{0} \right)$ is provided by (\ref{eq:T_func}) and
\begin{eqnarray}
\mathcal{S}\left( r_\mathsf{out}, r_{\mathsf{0}} \right) \hspace{-0.2cm} & = & \hspace{-0.2cm} \int
f_r(r)  \left( 1 -  p  +
\frac{ p r^{\alpha} }{\beta  +  r^{\alpha} } \right)^2 \,{\rm d} r \nonumber
\end{eqnarray}
\begin{eqnarray}
\hspace{-0.2cm} & = & \hspace{-0.2cm}
\left(1-p \right)^2 + \frac{4 r_{\mathsf{0}}^2}{|{\mathcal A}|} \left(1-p \right)p \left[ \frac{\pi}{\beta} \Psi \left( \frac{r_\mathsf{out}-r_{\mathsf{0}}}{r_{\mathsf{0}}} \right) \right. \nonumber
\end{eqnarray}
\begin{eqnarray}
\hspace{-0.2cm} &  & \hspace{-0.2cm}
\left.
+ \mathcal{Z}_1\left( r_\mathsf{out} , r_{\mathsf{0}} \right) \right] + \frac{2 r_{\mathsf{0}}^2 p^2}{|{\mathcal A}|} \left[ \pi \mathcal{G}\left( \frac{r_\mathsf{out} - r_{\mathsf{0}}}{r_{\mathsf{0}}} \right)  \right. \nonumber \\
\hspace{-0.2cm} &  & \hspace{-0.2cm}
+ \mathcal{Z}_2 \left( r_\mathsf{out} , r_{\mathsf{0}} \right) \bigg] \label{integral_S(x,y)}
\end{eqnarray}
where $\Psi \left(x \right)$ is given by (\ref{eq:Phi1}), $\mathcal{Z}_i \left( x, y \right)$ is given by (\ref{eq:Z_i}) and
\begin{eqnarray}
\mathcal{G}\left( x \right)
& = &
\int_{ 0 }^{ x }
\frac{r^{2 \alpha+1} }{\left(\beta + r^{\alpha}\right)^2 } \,{\rm d} r . \label{eq:G(x)}
\end{eqnarray}
By performing the change of variable $\displaystyle r = y  \nu^{\frac{1}{\alpha}}$, 
\begin{eqnarray}
\mathcal{G}\left( x \right)
 \hspace{-0.05cm} &= &\hspace{-0.05cm} \frac{x^{2\alpha +2}}{\left( 2+2\alpha\right) \beta^2 } \nonumber \\ && \times
  _{2}F_{1}\hspace{-0.05cm}
\left(\hspace{-0.05cm} \left[ 2, \hspace{-0.05cm} \frac{2+2\alpha}{\alpha}\hspace{-0.05cm} \right]\hspace{-0.05cm};\hspace{-0.05cm} \frac{2+3\alpha}{\alpha}\hspace{-0.05cm};\hspace{-0.05cm}-\frac{x^\alpha}{\beta} \right).
  \label{integral_G(x)_2}
\end{eqnarray}

By substituting (\ref{eq:T_func}), (\ref{integral_S(x,y)}) and (\ref{integral_G(x)_2}),  in (\ref{cdfwithintegral_square}), and by performing few algebraic manipulations, (\ref{eq:SecondMomentFinal}) is obtained.

\section{} \label{App:C}

This appendix provides details leading to (\ref{eq:JointFirstMomentFinal}). The spatially averaged joint first moment of the outage probability is derived from (\ref{eq:JointFirstMoment}). If polar coordinates are used to evaluate the normalized distances between each interferer and the $j^{th}$ receiver for each snapshot of the network, then
\begin{eqnarray}
\boldsymbol r\left(\rho,\phi, \theta \right) \hspace{-0.3cm} &=& \hspace{-0.3cm} \frac{\left| \left|r_\mathsf{0} \exp\left( j  \theta  \right) + r_\mathsf{out} \sqrt{\rho} \exp\left( j  \phi \right)\right|\right|}{r_\mathsf{0}}
\end{eqnarray}
where $\rho \sim \mathcal{U}\left(0,1 \right)$ and $\phi \sim \mathcal{U} \left(0,2 \pi \right)$. Since the $M$ interferers are i.u.d. over $\mathcal A$, it yields
(\ref{eq:JointFirstMomentFinal_Derivations_1}), where
\begin{eqnarray}
 R\left(\theta\right) \hspace{-0.3cm} &=& \hspace{-0.3cm} \int \hspace{-0.3cm} \int
\frac{r^{\alpha}\left(\rho,\phi, \theta \right) }{\beta + r^{\alpha} \left(\rho,\phi, \theta \right) } f_{\rho}\left( \rho\right)  f_{\phi}\left( \phi\right) \,{\rm d} \rho \,{\rm d} \phi.
\end{eqnarray}
The first and second double integral in (\ref{eq:JointFirstMomentFinal_Derivations_1}) are equivalent, since the first moment of the outage probability is the same at the two receivers, which are equally distant from the reference transmitter, and they coincide with (\ref{eq:T_func}).
Performing the cross products and few other algebraic manipulations in the last term of (\ref{eq:JointFirstMomentFinal_Derivations_1}) and by using the functions given by (\ref{Ihyp}) and (\ref{eq:Z_i}) yields
\begin{eqnarray}
\mathcal{W}\left( r_\mathsf{out}, r_\mathsf{0},\theta \right)  \hspace{-0.3cm} &=& \hspace{-0.3cm}
\left(1 - p \right)^2 \hspace{-0.05cm} +  p^2 \mathcal{X} \left( \theta\right) +  \hspace{-0.05cm} \frac{4 r_\mathsf{0}^2}{|\mathcal{A}|} \hspace{-0.05cm} \left(1 \hspace{-0.05cm}- \hspace{-0.05cm} p \hspace{-0.05cm} \right)\hspace{-0.05cm} p \hspace{-0.05cm}  \nonumber \\
\hspace{-0.3cm} & & \hspace{-0.3cm}
 \times \hspace{-0.05cm} \left[\hspace{-0.05cm} \pi \Psi_1\hspace{-0.05cm} \left( \hspace{-0.05cm} \frac{r_\mathsf{out}-  r_\mathsf{0}}{r_\mathsf{0}} \hspace{-0.05cm} \right) \hspace{-0.05cm} \hspace{-0.05cm} + \hspace{-0.05cm} \hspace{-0.05cm} \mathcal{Z}_1 \hspace{-0.05cm} \left( \hspace{-0.05cm} r_\mathsf{out}, r_\mathsf{0} \hspace{-0.05cm} \right)\hspace{-0.05cm} \right]   \label{eq:W(x,y,z)}.
\end{eqnarray}
where
\begin{eqnarray}
\hspace{-0.25cm} \mathcal{X}\hspace{-0.05cm} \left(\hspace{-0.05cm} \theta\right) \hspace{-0.05cm} \hspace{-0.30cm}&=& \hspace{-0.30cm} \hspace{-0.05cm} \hspace{-0.05cm} \frac{1}{2\pi} \hspace{-0.05cm} \hspace{-0.05cm} \int_{0}^1 \hspace{-0.1cm} \hspace{-0.05cm} \hspace{-0.05cm} \int_{0}^{2\pi} \hspace{-0.05cm} \hspace{-0.05cm}
   \hspace{-0.1cm} \hspace{-0.1cm} \frac{r^{\alpha}\left(\rho,\phi, 0 \right)}{\beta \hspace{-0.05cm} + \hspace{-0.05cm} r^{\alpha} \hspace{-0.05cm} \left(\rho,\phi, 0 \right)} \frac{r^{\alpha}\left(\rho,\phi, \theta \right)}{\beta  \hspace{-0.05cm} + \hspace{-0.05cm} r^{\alpha}\hspace{-0.05cm} \left(\rho,\phi, \theta \right)}
  \hspace{-0.05cm}  \hspace{-0.05cm} \,{\rm d} \mathbf{\rho} \hspace{-0.05cm} \,{\rm d} \mathbf{\phi}.
   \label{eq:X(x,y,z)}
\end{eqnarray}

Finally, By substituting (\ref{eq:W(x,y,z)}), and (\ref{eq:X(x,y,z)}),  in (\ref{eq:JointFirstMomentFinal_Derivations_1}), and by performing few algebraic manipulations, (\ref{eq:JointFirstMomentFinal}) is obtained.


\section{} \label{App:D}

This appendix provides details leading to (\ref{eq:JointFirstMomentFinalTCP}).
A similar approach of that in Appendix \ref{App:A} yields  \vspace{-0.1cm}
\begin{eqnarray}
   \mathbb E_{\boldsymbol  r }\left[\epsilon_j\left( \boldsymbol r \right) | m \right]
\hspace{-0.1cm} = \hspace{-0.1cm} 1 - \exp\left( -\frac{\beta }{\mathsf{SNR}} \right) \hspace{-0.1cm} \left[ 1-p + p \mathcal{V} \left( \sigma \right)
    \right]^{m} \hspace{-0.1cm}
\label{eq:firstmomentTCP}
\end{eqnarray}
where
\vspace{-0.2cm}
\begin{eqnarray}
\mathcal{V} \left( \sigma \right) &=& \int_{0}^{\infty} \frac{r^\alpha}{\beta+r^{\alpha}} f_{r}\left( r|\nu\right) {\rm d} r. \label{eq:V(sigma)}
\end{eqnarray}
If $\nu=0$, by performing a Taylor series expansion of (\ref{pdf_tx_cluster}), (\ref{eq:V(sigma)}) can be evaluated as follows  \vspace{-0.1cm}
\begin{eqnarray}
\hspace{-0.3cm} \mathcal{V} \left( \sigma \right) \hspace{-0.2cm} &=& \hspace{-0.2cm} \sum_{i=0}^{\infty} \frac{\left( -1\right)^i}{\beta} \frac{1}{i!}\frac{1}{2^i \sigma^{2+2i}}
\frac{x^{\alpha+2+2i}}{2+\alpha+2i} \nonumber \\  \vspace{-0.2cm}
\hspace{-0.2cm} & & \hspace{-0.2cm} _{2}F_{1} \hspace{-0.05cm}
\left( \hspace{-0.05cm}
\left[ \hspace{-0.05cm} 1, \frac{2+2i}{\alpha} \hspace{-0.05cm}+\hspace{-0.05cm}1 \hspace{-0.05cm} \right] \hspace{-0.05cm}; \hspace{-0.05cm}
\frac{2+2i}{\alpha} \hspace{-0.05cm} +\hspace{-0.05cm}   2 ;\hspace{-0.05cm}
- \frac{x^\alpha}{\beta } \hspace{-0.05cm}
\right) \hspace{-0.05cm} \Big|_{0}^{\infty}
 \label{eq:V(sigma2)}
\end{eqnarray}
where the series in (\ref{eq:V(sigma2)}) can be truncated after few terms. The last step is to uncondition over the number of interferers $m$, knowing that this follows a Poisson distribution. To do this, substitute (\ref{pmf_Poisson}), and (\ref{eq:V(sigma)}) in (\ref{eq:JointFirstMomentAveraged}) with $x=0$, and $y=1$. By applying the power-series representation of the exponential function, (\ref{eq:JointFirstMomentFinalTCP}) is finally obtained.


\vspace{-0.2cm}

\section{} \label{App:E}

This appendix provides details leading to (\ref{eq:SecondMomentFinal}).
Initially, the spatially averaged second moment is found conditioned on the number of interferers $m$ within the cluster located at distance $\nu$ from the center. This is obtained by substituting (\ref{Equation:RayleighConditional}) and \ref{pdf_tx_cluster} in (\ref{eq:SecondMoment}). Since the $m$ interferers are i.u.d., and after few algebraic manipulations,  it yields (\ref{cdfwithintegral_square_TCP}), where the function $\mathcal{V} \left( \sigma \right)$ is provided by (\ref{eq:V(sigma)}) and \vspace{-0.2cm}
\begin{eqnarray}
\mathcal{Q}\left( \sigma \right) \hspace{-0.2cm} & = & \hspace{-0.2cm} \int_0^{\infty}
f_r(r|v)  \left( 1 -  p  +
\frac{ p r^{\alpha} }{\beta  +  r^{\alpha} } \right)^2 \,{\rm d} r \nonumber  \\ \vspace{-0.2cm}
\hspace{-0.2cm} & = & \hspace{-0.2cm}
\left(1-p \right)^2 + 2 \left(1-p \right) p \mathcal{V} \left( \sigma \right) + p^2 \mathcal{C} \left( \sigma \right) \label{integral_Q(sigma)}
\end{eqnarray}
where $\mathcal{V} \left( \sigma \right)$ is given by (\ref{eq:V(sigma)}) and  \vspace{-0.05cm}
\begin{eqnarray}
\mathcal{V} \left( \sigma \right) &=& \int_{0}^{\infty} \frac{r^{2 \alpha}}{\left(\beta+r^{\alpha}\right)^2} f_{r}\left( r|\nu\right) {\rm d} r. \label{eq:C(sigma)}
\end{eqnarray}
If $\nu=0$, similar to Appendix \ref{App:D} by performing a Taylor series expansion of (\ref{pdf_tx_cluster}), (\ref{eq:C(sigma)}) can be evaluated as follows
\begin{eqnarray}
\hspace{-0.3cm} \mathcal{V} \left( \sigma \right) \hspace{-0.3cm} &=& \hspace{-0.3cm} \sum_{i=0}^{\infty} \frac{\left( -1\right)^i}{\beta} \frac{1}{i!}\frac{1}{2^i \sigma^{2+2i}}
\frac{x^{2 \alpha+2+2i}}{\left(2+2 \alpha+2i\right) \beta^2} \nonumber \\  \vspace{-0.2cm}
\hspace{-0.3cm} \hspace{-0.3cm} & & \hspace{-0.3cm} _{2}F_{1} \hspace{-0.05cm}
\left(\hspace{-0.05cm}
\left[ \hspace{-0.05cm} 2, \frac{2+2i}{\alpha} \hspace{-0.05cm} + \hspace{-0.05cm} 2 \right]\hspace{-0.05cm};\hspace{-0.05cm}
\frac{2+2i}{\alpha} \hspace{-0.05cm}+\hspace{-0.05cm}  3;
- \frac{x^\alpha}{\beta }\hspace{-0.05cm}
\right) \hspace{-0.05cm} \Big|_{0}^{\infty} \label{eq:C(sigma)Final}
\end{eqnarray}
where the series in (\ref{eq:C(sigma)Final}) can be truncated after few terms. Finally, substitute (\ref{pmf_Poisson}), and (\ref{cdfwithintegral_square_TCP}) in (\ref{eq:JointFirstMomentAveraged}) and set $x=0$, and $y=2$. By applying the power-series representation of the exponential function, (\ref{eq:JointSecondMomentFinalTCP}) is finally obtained.

\section{} \label{App:F}
This appendix provides details leading to (\ref{eq:JointFirstMomentFinalTCP2}). The spatially averaged joint first moment of the outage probability is derived from (\ref{eq:JointFirstMoment}). By first conditioning over the number of interferers, and by using a similar approach to that used in Appendix \ref{App:C}, it yields (\ref{eq:JointFirstMomentFinal_Derivations_1}), where

\begin{eqnarray}
\hspace{-0.4cm} \boldsymbol r\left(\rho,\phi, \theta \right) \hspace{-0.35cm} &=& \hspace{-0.35cm} \left| \left|r_\mathsf{0} \hspace{-0.05cm}\exp \hspace{-0.05cm} \left( \hspace{-0.05cm} j  \theta  \right)\hspace{-0.1cm} + \hspace{-0.1cm}\sqrt{\hspace{-0.05cm} - \hspace{-0.05cm} 2 \sigma^2 \hspace{-0.05cm}  \log\hspace{-0.05cm} \left(\hspace{-0.05cm}\rho\hspace{-0.05cm} \right)\hspace{-0.05cm} } \exp\hspace{-0.05cm}\left( \hspace{-0.05cm} j \phi \hspace{-0.05cm}\right)\hspace{-0.05cm}  \right|\right| \hspace{-0.05cm} /r_\mathsf{0} \label{eq:r}
\end{eqnarray}
where $\rho \sim \mathcal{U}\left(0,1 \right)$ and $\phi \sim \mathcal{U} \left(0,2 \pi \right)$.

The first and second double integral in (\ref{eq:JointFirstMomentFinal_Derivations_1}) are again equivalent, and they coincide with (\ref{eq:V(sigma)}). By performing the cross products in the last term of (\ref{eq:JointFirstMomentFinal_Derivations_1}), \vspace{-0.05cm}
\begin{eqnarray}
\mathcal{D}\left( \sigma,\theta \right)  \hspace{-0.3cm} &=& \hspace{-0.3cm}
\left(1 - p \right)^2 +   2 \left(1 - p \right) p \mathcal{V} \left( \sigma \right)+  p^2 \mathcal{X} \left( \theta\right) \label{integral_D(sigma)}.
\end{eqnarray}
where the function $\mathcal{X}\left(\theta\right)$ is given by (\ref{eq:X(x,y,z)}) once
(\ref{eq:r}) is substituted in it. Finally, in order to marginalize over the number of interferers, substitute (\ref{pmf_Poisson}), and (\ref{eq:JointFirstMomentFinal_Derivations_1}) in (\ref{eq:JointFirstMomentAveraged}) with $x=1$, and $y=1$. By applying the power-series representation of the exponential function, (\ref{eq:JointFirstMomentFinalTCP2}) is finally obtained.

\balance

\bibliographystyle{ieeetr}
\bibliography{CorrelationRef}

\begin{thebibliography}{10}

\bibitem{baccelli:2006}
F.~Baccelli, B.~Blaszczyszyn, and P.~Muhlethaler, ``An aloha protocol for
  multihop mobile wireless networks,'' {\em IEEE Trans. Inform. Theory},
  vol.~52, pp.~421--436, February 2006.

\bibitem{ganti:2009b}
M.~Haenggi and R.~K. Ganti, {\em Interference in Large Wireless Networks}.
\newblock Paris: Now, 2009.

\bibitem{haenggi:2012b}
M.~Haenggi, {\em Stochastic Geometry for Wireless Networks}.
\newblock Cambridge University Press, 2012.

\bibitem{torrieri:2012}
D.~Torrieri and M.~C. Valenti, ``The outage probability of a finite ad hoc
  network in {Nakagami} fading,'' {\em IEEE Trans. Commun.}, vol.~60,
  pp.~3509--3518, Nov. 2012.

\bibitem{valenti:2014}
M.~C. Valenti, D.~Torrieri, and S.~Talarico, ``A direct approach to computing
  spatially averaged outage probability,'' {\em IEEE Commun. Letters}, vol.~18,
  pp.~1103--1106, July 2014.

\bibitem{guo:2014}
J.~Guo, S.~Durrani, and X.~Zhou, ``Outage probability in arbitrarily-shaped
  finite wireless networks,'' {\em IEEE Trans. Commun.}, vol.~62, pp.~699--712,
  February 2014.

\bibitem{ganti:2009}
R.~K. Ganti and M.~Haenggi, ``Spatial and temporal correlation of the
  interference in {ALOHA} ad hoc networks,'' {\em IEEE Commun. Letters},
  vol.~13, pp.~631--633, September 2009.

\bibitem{haenggi:2012}
M.~Haenggi, ``Diversity loss due to intereference correlation,'' {\em IEEE
  Commun. Letters}, vol.~16, pp.~1600--1603, October 2012.

\bibitem{schilcher:2012}
U.~Schilcher, C.~Bettstetter, and G.~Brandner, ``Temporal correlation of
  interference in wireless networks with {Rayleigh} block fading,'' {\em IEEE
  Trans. Mobile Computing}, vol.~11, pp.~2109--2120, December 2012.

\bibitem{haenggi:2013d}
M.~Haenggi and R.~Smarandache, ``Diversity polynomials for the analysis of
  temporal correlations in wireless networks,'' {\em IEEE Trans. Wireless
  Comm.}, vol.~12, pp.~5940--5951, October 2013.

\bibitem{haenggi:2013}
M.~Haenggi, ``The local delay in {Poisson} networks,'' {\em IEEE Trans. Inform.
  Theory}, vol.~59, p.~1788–1802, March 2013.

\bibitem{zhong:2014}
Y.~Zhong, W.~Zhang, and M.~Haenggi, ``Managing interference correlation through
  random medium access,'' {\em IEEE Trans. Wireless Comm.}, vol.~13,
  pp.~928--941, February 2014.

\bibitem{gong:2014}
Z.~Gong and M.~Haenggi, ``Interference and outage in mobile random networks:
  Expectation, distribution, and correlation,'' {\em IEEE Trans. Mobile
  Computing}, vol.~13, pp.~337--349, February 2014.

\bibitem{tanbourgi:2014}
R.~Tanbourgi, H.~S. Dhillon, J.~G. Andrews, and F.~K. Jondral, ``Effect of
  spatial interference correlation on the performance of maximum ratio
  combining,'' {\em IEEE Trans. Wireless Comm.}, vol.~13, pp.~3307--3316, April
  2014.

\bibitem{wen:2015}
J.~Wen, M.~Sheng, B.~Liang, X.~Wang, Y.~Zhang, and J.~Li, ``Correlations of
  interference and link successes in heterogeneous cellular networks,'' in {\em
  Proc. IEEE Global Telecommun. Conf. (GLOBECOM)}, (San Diego, California),
  December 2015.

\bibitem{Schilcher:2016}
U.~Schilcher, S.~Toumpis, M.~Haenggi, A.~Crismani, G.~Brandner, and
  C.~Bettstetter, ``Interference functionals in poisson networks,'' {\em IEEE
  Trans. Inform. Theory}, vol.~62, p.~370–383, January 2016.

\bibitem{wen:2016}
J.~Wen, M.~Sheng, K.~Huang, and J.~Li, ``Analisys of interference correlation
  in non-poisson networks,'' in {\em Proc. IEEE Global Telecommun. Conf.
  (GLOBECOM)}, (Washington, DC), December 2016.

\bibitem{Ali:2017}
K.~S. Ali, H.~ElSawy, M.~Haenggi, and M.-S. Alouini, ``The effect of spatial
  interference correlation and jamming on secrecy in cellular networks,'' {\em
  IEEE Wireless Commun. Letters}, vol.~6, pp.~530--533, August 2017.

\bibitem{Krishnan:2017a}
S.~Krishnan and H.~S. Dhillon, ``Exact characterization of spatio-temporal
  joint coverage probability in cellular networks,'' in {\em Proc. IEEE
  Wireless Commun. and Networking Conf.}, (San Francisco, CA), March 2017.

\bibitem{Krishnan:2017}
S.~Krishnan and H.~S. Dhillon, ``Effect of user mobility on the performance of
  device-to-device networks with distributed caching,'' {\em IEEE Wireless
  Commun. Letters}, vol.~6, pp.~194--197, January 2017.

\bibitem{Rajanna:2017}
A.~Rajanna and M.~Haenggi, ``Downlink coordinated joint transmission for mutual
  information accumulation,'' {\em IEEE Wireless Commun. Letters}, vol.~6,
  pp.~198--201, April 2017.

\bibitem{ganti2}
R.~K. Ganti and M.~Haenggi, ``Interference and outage in clustered wireless ad
  hoc networks,'' {\em IEEE Trans. Inform. Theory}, vol.~55, pp.~4067--4086,
  September 2009.

\bibitem{Afshang:2016}
M.~Afshang, H.~S. Dhillon, and P.~H.~J. Chong, ``Modeling and performance
  analysis of clustered device-to-device networks,'' {\em IEEE Trans. Wireless
  Comm.}, vol.~15, pp.~4957--4972, July 2016.

\bibitem{khalid:2013}
Z.~Khalid and S.~Durrani, ``Distance distributions in regular polygons,'' {\em
  IEEE Trans. Veh. Tech.}, vol.~62, pp.~2363--2368, June 2013.

\end{thebibliography}

\balance

\end{document}